\documentclass[aps,twocolumn,prd,nofootinbib]{revtex4}
\usepackage{amsmath}
\usepackage{graphicx}
\usepackage{dcolumn}
\usepackage{bm}
\usepackage{amssymb}
\usepackage{latexsym}

\def\be{\begin{equation}}
\def\ee{\end{equation}}
\def\bea{\begin{eqnarray}}
\def\eea{\end{eqnarray}}

\begin{document}

\title{Supernova Constraints on Models of Neutrino Dark Energy}

\author{Hong Li$^{a}$, Bo Feng$^{b}$, Jun-Qing Xia$^{a}$ and Xinmin Zhang$^{a}$}
\affiliation{$^{a}$Institute of High Energy Physics, Chinese
Academy of Sciences, P.O. Box 918-4, Beijing 100049, P. R. China}
\affiliation{$^{b}$ National Astronomical Observatories, Chinese
Academy of Sciences, 20A Datun Road, Beijing 100012, P. R. China}

\begin{abstract}
In this paper we use the recently released Type Ia Supernova (SNIa)
data to constrain the interactions between the neutrinos and the
dark energy scalar fields. In the analysis we take the dark energy
scalars to be either Quintessence-like or Phantom-like. Our results
show the data mildly favor a model where the neutrinos couple to a
phantom-like dark energy scalar, which implies the equation of state
of the coupled system behaves like Quintom scenario in the sense of
parameter degeneracy. We find future observations like SNAP are
potentially promising to measure the couplings between neutrino and
dark energy.

\end{abstract}

\maketitle

\hskip 1.6cm PACS number(s): 98.80.Es \vskip 0.4cm

\section{Introduction}

Astronomical observations of the Type Ia Supernova (SNIa), Cosmic
Microwave Background Radiation(CMB) and the Large Scale
Structure(LSS) strongly support for a concordance model of cosmology
where the universe is flat and composed of around seventy percent of
dark energy(DE)\cite{Riess04,Spergel03,0310723}. The simplest
candidate for dark energy seems to be a remnant small cosmological
constant. However, many physicists are attracted by the idea that
dark energy is due to a dynamical component, such as a canonical
scalar field $Q$, named {\it Quintessence}\cite{quintessence}. The
recent fits to the SNIa data and CMB etc in the literature find that
the behavior of dark energy is to great extent in consistency with a
cosmological constant, however the dynamical dark energy scenarios
are generally not ruled out and in fact one class of models with an
equation of state(EOS) transiting from below -1 to above -1 as the
redshift increases, {\it Quintom}\cite{FWZ,Qx} is mildly
favored\cite{quintom}. Being a dynamical component, the scalar field
of dark energy is expected to interact with the ordinary matters. If
these interactions  exist, it will open up the possibility of
detecting the dark energy non-gravitationally.

Recently there have been a lot of interests in the literature in
studying the possible connections between the neutrinos and the dark
energy.\cite{paper30,paper11,paper15,paper31,paper32,paper33,paper34,
paper35,paper36,paper37,paper38,paper39,paper40,grb,BFLZ,FR03,twofield}
 There seem to be at least two reasons which
motivate these studies: 1) the dark energy scale $\sim 10^{-3}$ ev
is smaller than the energy scales in particle physics, but
interestingly is comparable to the neutrino masses; 2) in
Quintessence-like models of dark energy $m_Q \sim 10^{-33}$ eV,
which surprisingly is also connected to the neutrino masses {\it
via} a see-saw formula $m_Q \sim {m_\nu^2 / m_{pl}}$ with $m_{pl}$
the planck mass.

Is there really any connections between the neutrinos and dark
energy? Given the arguments above it is quite interesting to make
such a speculation on this connection. If yes, however in terms of
the language of the particle physics it requires the existence of
new dynamics and new interactions between the neutrinos and the
dark energy sector.

In general for the models of neutrino dark energy or interacting
dark energy, the lagrangian can be written as
\begin{eqnarray}
{\cal L}= {\cal L}^{SM}_{\nu} +  {\cal L}_{\phi} + {\cal L}_{int},
\end{eqnarray}
where ${\cal L}^{SM}_{\nu}$ is the lagrangian of the standard model
(SM) describing
the physics of the left-handed neutrinos, ${\cal L}_{\phi}$ is for
the dynamical scalar such as Quintessence. ${\cal L}_{int}$ in (1) is the
sector which mediates the interaction between the dark energy
scalar and the  neutrinos.

At energy much below the electroweak scale, the relevant lagrangian for
the neutrino dark energy is given by
\begin{equation}
{\cal L}={\cal L}_\nu + {\cal L}_\phi
+M_{\nu}(\phi)\bar{\nu}{\nu}\ ,
\end{equation}
where ${\cal L}_\nu$ is the kinetic term of the neutrinos. For the
dark energy scalar part of ${\cal L}_\phi$, two types of models have often
been considered with one being the
quintessence-like and another phantom-like\cite{phantom}. Thus we
introduce a factor $F$ in the front of the kinetic term of the
scalar
 \begin{equation}\label{QFL}
  {\cal L}_\phi=\frac{F}{2}\partial_\mu
  \phi\partial^\mu\phi-V(\phi).
  \end{equation}
When $F=1$ it corresponds to quintessence-like and $F=-1$ for
phantom-like. $V(\phi)$ is the potential of the scalar field.

The last term of Eq.(2) is the scalar field dependent mass of the
neutrinos which characterizes the interaction between the
neutrinos and the dark energy scalar. In the standard model of
particle physics, the neutrino masses can be described by a
dimension-5 operator
  \begin{equation}
  L_{\not L}=\frac{2}{f}l_{L}l_{L}HH+h.c,
  \end{equation}
  where $f$ is a scale of new physics beyond the Standard Model
  which generates the $B-L$ violations, $l_{L},  H$ are the
  left-handed lepton and Higgs doublets respectively. When the
  Higgs field gets a vacuum expectation value $<H> \sim v$, the
  left-handed neutrino receives a majorana mass
  $m_{\nu} \sim \frac{v^{2}}{f}$. In Ref.\cite{paper31} we considered
   an interaction between the
neutrinos and the Quintessence $\phi$
 \begin{equation}
 \beta \frac { \phi }{M_{pl}} \frac{2 }{f} l_{L}l_{L}
HH+ h.c  ,
 \end{equation}
 where $\beta $ is the coefficient which characterizes the strength of the
Quintessence interacting with the neutrinos. In this scenario the
neutrino masses vary during the evolution of the universe and we
have shown that the neutrino mass limits imposed by the
baryogenesis are modified.

The dim-5 operator above is not renormalizable, which in principle
can be generated by integrating out the heavy particles. For
example, in the model of the minimal see-saw mechanism for the
neutrino masses,
\begin{equation}
L=h_{ij}\bar{l}_{Li}N_{Rj}H+\frac{1}{2}M_{ij}\bar{N}^{c}_{Ri}N_{Rj}+h.c.
 \end{equation}
 where $ M_{ij}$ is the mass matrix of
the right-handed neutrinos and the Dirac masses of the neutrinos are given
by $m_{D}\equiv h_{ij} <H> $. Integrating out the heavy
right-handed neutrinos one will generate a dim-5 operator, however as
pointed out in Ref.\cite{paper31} to have the light neutrino
masses varied there are various possibilities, such as by coupling
the Quintessence field to either the Dirac masses or the majorana
masses of the right-handed neutrinos or both.

In Ref.\cite{paper33} we have specifically proposed a model of
mass varying right-handed neutrinos. In this model the
right-handed neutrino masses $M_i$ are assumed to be a function of
the Quintessence scalar $M_i(\phi)=\overline{M}_i e^{\beta
\frac{\phi}{M_{pl}}}$. Integrating out the right-handed neutrino
will generate a dimension-5 operator, but for this case the light
neutrino masses will vary in the following way
\begin{equation}
 e^{-\beta
\frac{\phi}{M_{pl}}} \frac{2 }{f} l_{L}l_{L} HH+ h.c  .
 \end{equation}
With mass varying right-handed neutrinos given above we have in
\cite{paper33} studied in detail its implication in thermal
leptogenesis. In Ref.\cite{grb} we have studied the possibility of
detecting the time variation of the neutrino masses with Short Gamma
Ray Burst. In Ref. \cite{BFLZ} we have discussed the implications of
the mass varying neutrinos in the cosmological evolution of the
Universe. And in Ref.\cite{twofield} we argued that neutrinos
coupled to Phantom scalar can provide a scenario of dark energy with
the equation of state crossing the cosmological constant boundary of
-1. In this paper we will use the recently released  SNIa data to
constrain the couplings between the neutrinos and the dark energy
scalar. We will show that the current data mildly favor the model
where the Phantom-like scalar couples to the neutrinos. This paper
is organized as follows: in section II we will present the
formulation and the results on the constraint on the equation of
state of the coupled system; In section III we study the constraints
on the couplings of the neutrinos to the scalar fields; Section IV
is our conclusion.

\section{Formulation and the constraint on the equation of state}

In this section we will study in detail the scenario of the
coupled system of neutrino and scalar field described by the
lagrangian in (2). The equation of motion of the scalar field
$\phi$ is given by \be\label{EM}
F(\ddot{\phi}+3H\dot{\phi})+\frac{dV}{d\phi}+\frac{dV_I}{d\phi}=0\
, \ee where \be \label{sour}
\frac{dV_I}{d\phi}=\frac{dM_{\nu}}{d\phi}n_{\nu}\left\langle\frac{M_{\nu}}{E}\right\rangle
\ee is the source term by the interaction between the neutrinos
and dark energy, with $n_{\nu}$ and $E$ being the number density
and energy of the neutrinos respectively and $\langle \rangle$
indicating the thermal average. For relativistic neutrinos, the
term $\frac{dV_I}{d\phi}$ is greatly suppressed and the neutrinos
and dark energy decouple. For non-relativistic neutrinos, the
effective potential of the system is given by
$V_{eff}(\phi)=V(\phi)+n_{\nu}M_{\nu}(\phi)$. In the following,
for the simplicity we take the $\phi$-depending neutrino mass
$M_{\nu}(\phi)$ as $M_{\nu}$.

For this coupled system, the energy for each component does not
conserve while the total number of neutrinos is constant. It is easy
to get that the energy density of neutrinos follows \be
\label{ronu}\dot{\rho}_\nu+3H\rho_\nu=n_{\nu}\dot{M_{\nu}}. \ee The
conservation of energy-momentum tensor of the whole system gives
that the fluid equation of the scalar field $\phi$ is
\be\label{rphi}
\dot{\rho}_\phi+3H\rho_\phi(1+w_\phi)=-n_{\nu}\dot{M_{\nu}}\ , \ee
with the corresponding equation of state $w_{\phi}\equiv
P_{\phi}/\rho_{\phi}=\frac{\frac{1}{2}F
\dot{\phi}^2-V(\phi)}{\frac{1}{2}F \dot{\phi}^2+V(\phi)}$, which
stands for the uncoupled equation of state and F has the same
convention as that in Eq. (\ref{QFL}).

From Eqs. (\ref{ronu}) and (\ref{rphi}) one can easily obtain the
equation of state of the whole coupled system \be \label{whole w}
w=\frac{\Omega_{\phi}}{\Omega_{\nu}+\Omega_{\phi}} w_{\phi}. \ee In
the following, we will use the recently released  SNIa
data\cite{Riess04} to constrain this class of models. For the low
redshift range covered by the geometric constraints of SNIa the
equation of state of the coupled system can be simply parameterized
as $w = w_0 + w_1 z$. In the left panel of Fig. 1 we plot the
parameter space of $( w_0, w_1 )$ constrained by the SNIa data. One
can see the best fitting values are around $w_0=-1.75$ and
$w_1=1.9$, which gives rise to an evolving EOS and transiting from
below -1 to above -1 as the redshift increases as shown in the right
panel of Fig.1. In our case for a constant equation of state the
best fitting value corresponds to
$w=-1.45$. 
On the current SNIa fittings we have assumed the matter energy
density fraction to be in the range $0.2<\Omega_m<0.4$, which is
somewhat optimistic. This will inevitably lead to some bias but does
not affect the physical picture of this paper. On the degeneracies
between $\Omega_m$ and dark energy EOS see e.g.
\cite{CMMS04,FWZ,0405218,Tao04}. Generically one needs to add some
priors or combined data analysis to extract the behaviors of
dynamical dark energy\cite{Riess04,quintom}. A thorough combined
observational constraints on the couplings needs the modification of
the Boltzman code and investigations on the full set of parameter
space, which is extremely time-consuming in computation with the
massive neutrinos (for a preliminary model-dependent study see
\cite{BBMT05}). In this paper we assume a flat prior of the
Universe, i.e. $\Omega_k=0$.

\begin{figure}[htbp]
\begin{center}
\includegraphics[scale=0.8]{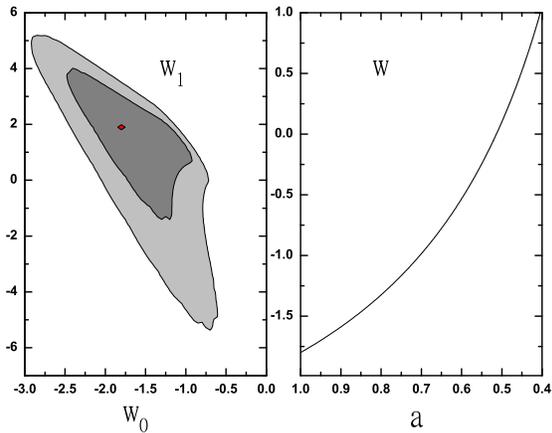}
\caption{The left panel: 2 $\sigma$ SNIa limit on the Dark Energy
model whose equation of state is $w = w_0 + w_1 z$ with prior
$0.2<\Omega_m<0.4$; The right panel: the evolution of $w$ as
function of scale factor $a$ with the center value given by the
left panel. }
\end{center}
\end{figure}

From equation $(\ref{whole w})$, one can see that to have $w$
transit from below $-1$ to above $-1$ as the redshift increases we
need $w_{\phi}<-1$ which implies the scalar field to be a
Phantom-like. In the next section we will use the SNIa data to
constrain the coupling of the neutrinos to the scalar field. We
will point out that the interaction between the neutrinos and the
phantom field is crucial to the Quintom scenario of dark energy.

\section{Constraint on the coupling of neutrinos to the scalar field}

Current experiments have put somewhat stringent constraints on the
conventional neutrino mass. Atmosphere neutrino oscillations show
that there is at least one species of neutrino with mass $\gtrsim
0.05$ eV\cite{SuperK}. Measurements of LSS matter power spectrum
are very sensitive to the total mass of neutrino and a combined
analysis from WMAP\cite{Spergel03} and SDSS\cite{0310723} gives an
upper bound $\Sigma m_{\nu} <1.7$ eV at $95 \%$ C.L. A recent
analysis has optimistically constrained the upper bound to be 0.42
eV. Cosmological combined constraints on  $\Sigma m_{\nu}$ are
faced with many parameter degeneracies and a recent interesting
study\cite{Hannestad} has shown the degeneracy between dark energy
EOS and neutrino mass, which can loosen the upper bound of
neutrino mass. When neutrino mass is varying with time the upper
bound on current $\Sigma m_{\nu}$ would be reasonably relaxed and
Ref.\cite{paper36} has quoted the value to be around 3 eV.
Neutrino density fraction is related to the neutrino mass by \be
\Omega_{\nu} h^2 = \frac{\Sigma m_{\nu}}{92.5 eV} ,\label{sigmamu}
\ee where h is Hubble parameter in units of 100 km s$^{-1}$
Mpc$^{-1}$. For a quantitative study below on the couplings
between neutrino and dark energy scalar we will take
$\Omega_{\nu}$ to be around $0.01$, $0.03$ and $0.06$ as specific
examples.

In the presence of the interaction between the dark energy scalar
and the neutrinos, the neutrino mass will vary as the universe
expands. In general different couplings give rise to different
behaviors of the mass variation. In the following we will first
give a model independent analysis and then go to the detail
models.

From Eqs.(\ref{ronu}) and (\ref{rphi}) we can get \be\label{ronu1}
\rho_{\nu}=\rho_{\nu 0}a^{-3}\frac{M_{\nu}}{M_{\nu 0}}, \ee where
$M_{\nu 0}$ is the mass of neutrino at present, $a$ being the scale
factor and \be\label{rophi1} \rho_{\phi}=\rho_{\phi
0}a^{-3(1+w_{\phi})}(1-\frac{1}{\rho_{\phi 0}}\int_{M_{\nu
0}}^{M_{\nu}} n_{\nu}a^{3(1+w_{\phi})}d m_{\nu}). \ee

From the Eqs.(\ref{ronu1}) and (\ref{rophi1}) we eventually get
\begin{eqnarray}\label{rhototal}
\rho_{tot}=\rho_0[{\Omega_{m}}_0 a^{-3}+{\Omega_{\nu}}_0
a^{-3}\frac{M_{\nu}}{M_{\nu 0}}+{\Omega_{\phi}}_0 X(a)],
\end{eqnarray}
with $\rho_0$ being the current critical density and
\begin{equation}\label{xa}
X(a)\equiv  a^{-3(1+w_{\phi})}(1-\frac{1}{\rho_{\phi 0}}\int_{M_{\nu
0}}^{M_{\nu}} n_{\nu}a^{3(1+w_{\phi})}d m_{\nu}),
\end{equation}
 and the Friedman equation
\begin{equation}
H^2=H_0^2E(a)^2,\label{h}
\end{equation}
where \be\label{ea2} E(a)^2\equiv [{\Omega_{m}}_0
a^{-3}+{\Omega_{\nu}}_0 a^{-3}\frac{M_{\nu}}{M_{\nu
0}}+{\Omega_{\phi}}_0 X(a)]. \ee In a flat Universe, the observed
luminosity distance of SNIa can be expressed as
\begin{equation}
d_L(a)=\Gamma(a)c/(aH_0),
\end{equation}
 where $\Gamma(a)=\int_a^1 {a^{\prime}}^{-2}
da^{\prime}/E(a^{\prime})$ and $c$ is the speed of light. In the
considerations above we have assumed that $w_\phi$ is constant for
the simplicity of the discussions.

Now we move to the detail model. To have a model independent
analysis we first expand the neutrino mass in the powers of the
$\ln a$ with $a$ being the scale factor. Setting $u=\ln a$ we
have:
\begin{eqnarray}\label{double}
   M_{\nu}&=&M_{\nu 0}+M^{\prime}_{\nu}(0)u+\frac{1}{2}M^{\prime\prime}_{\nu}(0)u^2+...
     \nonumber\\
           &=&M_{\nu 0}(1+\frac{M^{\prime}(0)}{M_{\nu 0}}u+...).
    \end{eqnarray}
Defining $\beta\equiv
    \frac{M^{\prime}(0)}{M_{\nu 0}}$ and keeping the leading order in
$u$ one obtains
    \begin{equation}\label{mnu}
     M_{\nu} = M_{\nu 0}( 1+ \beta u ),
    \end{equation}
    where $M_{\nu 0}$ denotes the neutrino mass at the present time
and $\beta$ is the parameter characterizing the dependence of the
neutrino mass on $u$.  Substituting Eq. (\ref{mnu}) into Eq.
(\ref{rhototal}) and following Eq. (\ref{xa}), and combing the
conservation equation of the matter including the baryons and the
cold dark matter
\begin{equation}
\dot{\rho_{m}}+3H\rho_{m}=0.\label{eqm}
\end{equation}
we can get
\begin{eqnarray}
\rho_{tot}=\rho_0[{\Omega_{m}}_0 a^{-3}+{\Omega_{\nu}}_0
a^{-3}(1+\beta\ln a)+{\Omega_{\phi}}_0 X(a)],
\end{eqnarray}
with
\begin{equation}
X(a)\equiv
[1+\frac{\beta{\Omega_{\nu}}_0}{3w_{\phi}\Omega_{\phi_0}}(1-a^{3w_{\phi}})]a^{-3(1+w_{\phi})},
\end{equation}
 the Eq. (\ref{ea2}) now becomes
\be \label{p1} E(a)^2\equiv [{\Omega_{m}}_0
a^{-3}+{\Omega_{\nu}}_0 a^{-3}(1+\beta\ln a)+{\Omega_{\phi}}_0
X(a)]. \ee

\begin{figure}[htbp]
\begin{center}
\includegraphics[scale=0.8]{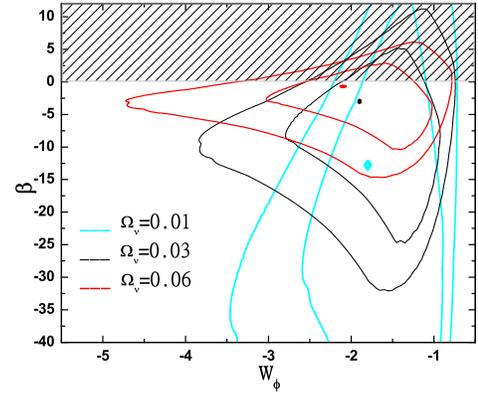}
\caption{ 2$\sigma$ SNIa limits on the coupled neutrino-DE model
with neutrino mass evolving as $M_{\nu} = M_{\nu 0}( 1+ \beta u )$,
where $u$ stands for $\ln a$. The dots inside show the best fit
values. The cyan(light dashed), black(solid) and red(dark dashed)
contours stand for $\Omega_{\nu}=0.01,0.03$ and 0.06 respectively.
The hatched area denotes where the neutrino mass might be negative
on large redshifts and physically forbidden. In the fittings we've
applied the uniform prior $0.2<\Omega_m<0.4$.\label{fig:lin} }
\end{center}
\end{figure}

\begin{figure}[htbp]
\begin{center}
\includegraphics[scale=0.8]{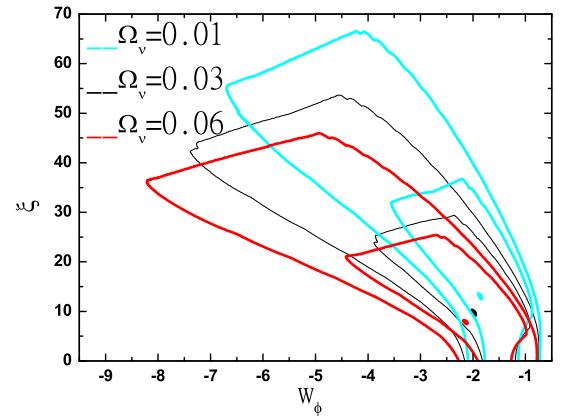}
\caption{ 2$\sigma$ SNIa limits on the coupled neutrino-DE model
with neutrino energy density evolving as
$\rho_{\nu}=\frac{\Omega_{\nu 0}}{\Omega_{\phi
0}}\rho_{\phi}a^{-\xi}$, where the subscript '0' stands for today's
value. The dots inside show the best fit values. The cyan(light
dashed), black(solid) and red(dark dashed) contours stand for
$\Omega_{\nu}=0.01,0.03$ and 0.06 respectively. In the fittings
we've applied the uniform prior $0.2<\Omega_m<0.4$.\label{fig:LC}}
\end{center}
\end{figure}

\begin{figure}[htbp]
\begin{center}
\includegraphics[scale=0.6]{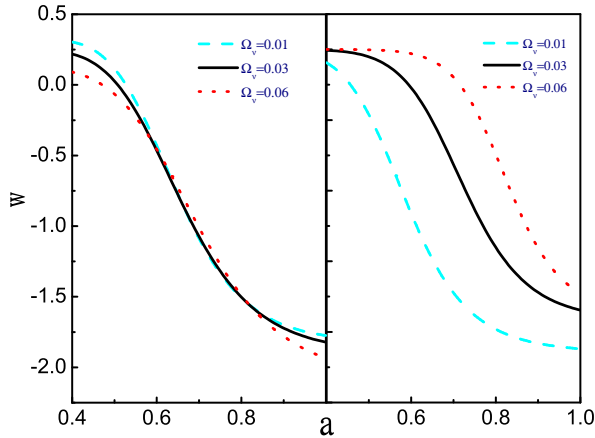}
\caption{ Effective EOS for the coupled neutrino-DE system for the
best-fit values of the two models shown in Fig.\ref{fig:lin} and
Fig.\ref{fig:LC}. The cyan(dashed), black(solid) and red(dotted)
lines stand for $\Omega_{\nu}=0.01,0.03$ and 0.06 respectively. The
left panel stands for the model with $M_{\nu} = M_{\nu 0}( 1+ \beta
u )$ and the right panel for $\rho_{\nu}=\frac{\Omega_{\nu
0}}{\Omega_{\phi 0}}\rho_{\phi}a^{-\xi}$.}
\end{center}
\end{figure}

With a uniform prior $0.2<\Omega_m<0.4$ in Fig. 2 we delineate
2$\sigma$ SNIa constraints on the parameter space $(\beta, w_\phi )$
of our model, where the SNIa data are taken from the Riess Gold
sample\cite{Riess04}. We can see that a non-vanishing $\beta$ is
preferred and the data mildly favor the model where the neutrinos
interact with the phantom-like scalar. The parameter space gets
better constrained with the increase of $\Omega_{\nu}$. Normally we
expect that the constraints of the parameter $\beta$ should be
$\mid\beta\mid < 1$ for the consideration of the small $\ln a$
expansion, but the current SNIa constraint is very weak which lead
to the $\ln a$ expansion of $M_{\nu}$ in Eq. (\ref{mnu}) loses its
significance in some sense as a large $|\beta|$ is not ruled out and
even somewhat favored. Moreover, the linear parametrization itself
is not well defined for a large but positive $\beta$, because the
neutrino becomes massless and even be with a negative mass on large
redshifts. In our Fig.2 the hatched denotes where $\beta>0$. We
should point out that in our numerical calculations we used no
approximations and just set Eq. (\ref{mnu}) as a parametrization.

Given the limitation mentioned above and the conventions in the
literature for the parametrization of the EOS of dark energy one may
consider different types of parametrization of the neutrino mass.
For an example we consider here another type of parametrization
which was firstly invoked for the study of dark energy coupled with
dark matter in solving the coincidence problem\cite{DL} \be
\frac{\rho_{\phi}}{\rho_{\nu}}=Aa^{\xi}, \label{m2} \ee where $A$
and $\xi$ are constant. $A$ is determined by
$A=\frac{\Omega_{\phi_0}}{\Omega_{\nu_0}}$, and the variation of the
neutrino mass is $d \ln M_{\nu}/ d \ln a =
-\frac{\Omega_{\phi_0}(\xi+3w_{\phi})}{\Omega_{\phi_0}+
\Omega_{\nu_0}a^{-\xi}}$. Correspondingly we now get \be\label{p2}
E(a)=(1-\Omega_{m_0})a^{-3}(\frac{\Omega_{\nu_0}+\Omega_{\phi_0}a^{\xi}}
{\Omega_{\nu_0}+\Omega_{\phi_0}})^{-\frac{3w_{\phi}}{\xi}}+\Omega_{m_0}a^{-3}.
\ee There would be interactions between neutrinos and the scalar
dark energy when $\xi\neq -3w_{\phi}$. $\xi=3,w_{\phi}=-1$ stands
for the conventional non-interacting $\Lambda$CDM cosmology and
$\xi=0$ for the "tracking" neutrino which would behave the same as
dark energy. As shown in Ref.\cite{BFLZ} it is not applicable that
neutrino and DE can enter the tracking regime as early as today,
since neutrino takes up a very small density fraction and dark
energy only comes to dominate the universe very recently. Hence we
will restrict our attentions on the parameter space where $\xi>0$.

In Fig.3 we delineate the  2$\sigma$ SNIa constraints on the
parameter space of $(\xi, w_\phi )$. The cyan(light dashed),
black(solid) and red(dark dashed) contours stand for
$\Omega_{\nu}=0.01,0.03$ and 0.06 respectively. We can see that
current SNIa constraints are very weak and the data favor mildly the
model where neutrinos interact with the phantom-like scalar. It is
noteworthy that here the best fitting values show $\xi+3w_{\phi}>0$,
which corresponds to an increase of neutrino mass with the redshift,
and the behavior is similar to our first example. We find for the
two parametrizations when the prior of $\Omega_m$ is relaxed the
parameter space would be constrained less stringently, but a
phantom-like DE coupling with neutrinos is still mildly favored.

More recently the authors of Ref.\cite{snls} made the distance
measurements to 71 high redshift type Ia supernovae discovered
during the first year of the 5-year Supernova Legacy Survey (SNLS).
SNLS will hopefully discover around 700 type Ia supernovas, which is
an intriguing ongoing project. When performing the current SNLS
constraints on the neutrino-DE coupling, we found that the current
SNLS data have not yet been as precise as the Riess Gold
sample\cite{Riess04} and the conclusion above has not changed.

As shown by Eq.$(\ref{whole w})$, for SNIa data only the preference
for a Quintom-like behavior of dark energy (where the equation of
state gets across -1) is fully degenerate with the preference of a
coupling between neutrinos and phantom-like dark
energy($w_{\phi}<-1$). In fact the fittings in Section II can also
be rephrased as current SNIa constraints on the linearly
parametrized EOS of single component of dark energy, as shown in
Ref.\cite{FWZ}. To give a more intuitive example we delineate in
Fig.4 the effective EOS of the full coupled system given by the best
fit values of Fig.\ref{fig:lin} and Fig.\ref{fig:LC}. The cyan(light
dashed), black(solid) and red(dark dashed) lines stand for
$\Omega_{\nu}=0.01,0.03$ and 0.06 respectively. The left panel
stands for the model with $M_{\nu} = M_{\nu 0}( 1+ \beta u )$ and
the right panel for $\rho_{\nu}=\frac{\Omega_{\nu 0}}{\Omega_{\phi
0}}\rho_{\phi}a^{-\xi}$. We can see they all show a Quintom-like
behavior. Actually from Eqs.(\ref{p1}) and (\ref{p2}) we can easily
work out the analytic forms of the effective EOS:
\begin{eqnarray}
w&=&-1-\frac{1}{3D} \{\Omega_{\nu_0}a^{-3}(-3+\beta-3\beta\ln a)
\nonumber\\
&+&\Omega_{\phi_0}[-3(B+1)(1+w_{\phi})a^{-3(1+w_{\phi})}
\nonumber\\
&+&3Ba^{-3}]\},
\end{eqnarray}
 where \be D=\Omega_{\nu_0}a^{-3}(1+\beta\ln a)+\Omega_{\phi_0}X(a),\ee
 and \be B=\frac{\beta\Omega_{\nu_0}}{3w_{\phi}\Omega_{\phi_0}}\ee
 for the coupling form of Eq.(\ref{m2}) and
 \be
w=w_{\phi}\frac{\Omega_{\phi_0}a^{\xi}}{\Omega_{\nu_0}+\Omega_{\phi_0}a^{\xi}},
\ee for the form of Eq.(\ref{p2}), which could easily get across
the cosmological constant boundary for the given best fit values
shown in Fig.\ref{fig:lin} and Fig.\ref{fig:LC}.

The projected satellite SNAP (Supernova / Acceleration Probe) would
be a space based telescope with a one square degree field of view
with 1 billion pixels. It aims to increase the discovery rate for
SNIa to about 2,000 per year\cite{snap}. In the following we will
use the simulated three-year SNAP data to investigate to what an
extent future SNAP would be able to detect or rule out the currently
mildly favored couplings between dark energy and neutrino. In
addition we assume a gaussian prior $\sigma_m=0.01$, it is close to
future Planck constraints\footnote{We found through our simulations
the constraints on neutrino-DE coupling would be weaker without the
prior on matter, where the necessity of the prior can be understood
from the following Figures 5 and 6. }.

The underlying fiducial model used here is the uncoupled
$\Lambda$CDM model with $w_{\phi}=-1,\Omega_{\phi}=0.7$ and $h=0.7$.
It corresponds to $\beta=0$ in the coupling form of Eq.(\ref{mnu})
and $\xi=3$ in the form of Eq.(\ref{m2}). The simulated SNIa data
distribution for each year is taken from Refs. \cite{kim, tao}. As
we consider 3-year data, the number for each bin will be improved by
around $3$ times. As for the error, we follow the ref. \cite{kim}
which takes the magnitude dispersion $0.15$ and the systematic error
$\sigma_{sys}(z)=0.02*z/1.7$, and the whole error for each data is
\begin{equation}
\sigma_{mag}(z_i)=\sqrt{\sigma_{sys}^2(z_i)+\frac{0.15^2}{n_i}},
\end{equation}
where $n_i$ is the number of supernova in the i'th redshift bin.

In Figure 5 we show the 3-year SNAP constraints on the coupled
neutrino-DE model given by Eq.(\ref{mnu}).  The red(light dashed),
blue(dark dashed) and green(solid) 2$\sigma$ contours stand for
$\Omega_{\nu}=0.01,0.03$ and 0.06 respectively. We find the coupling
parameter of $\beta$ is restricted to be of order unity and the
parametrization of $\ln a$ has made its sense when the neutrino
density fraction can be order of $0.01$.

Correspondingly in Figure 6 we delineate the 3-year SNAP constraints
on the coupled neutrino-DE model given by Eq.(\ref{p2}). The
red(dashed), blue(dash dotted) and green(solid) 2$\sigma$ contours
stand for $\Omega_{\nu}= 0.01,0.03$ and 0.06 respectively. The
behavior of this  parametrization is not symmetric around the
fiducial $\Lambda$CDM model, and the constraints on the parameters
are similar to that of the linear parametrization. Thus the future
projects like SNAP can improve the precision efficiently and is
potentially able to detect the couplings between neutrino and dark
energy.

\begin{figure}[htbp]
\begin{center}
\includegraphics[scale=0.65]{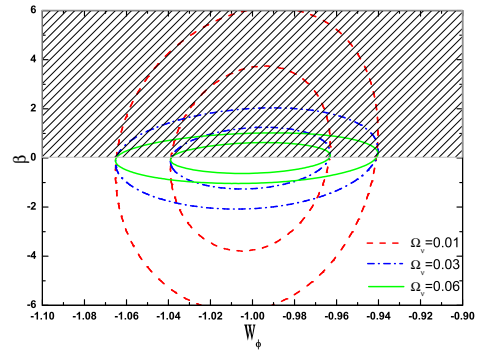}
\caption{Simulated 3-year SNAP constraints on the coupled
neutrino-DE model with neutrino mass evolving as $M_{\nu}(u) =
M_{\nu}(0)( 1+ \beta u )$, where $u$ stands for $\ln a$. The
red(dashed), blue(dash dotted) and green(solid) 2$\sigma$ contours
stand for $\Omega_{\nu}=0.01,0.03$ and 0.06 respectively. The
hatched area denotes where the neutrino mass might be negative on
large redshifts and physically forbidden. The underlying fiducial
model has been chosen as the uncoupled $\Lambda$CDM model with
$c=0,w_{\phi}=-1,\Omega_{\phi}=0.7$ and $h=0.7$. We take a gaussian
prior $\sigma_m=0.01$. \label{fig:SNAPLin} }
\end{center}
\end{figure}

\begin{figure}[htbp]
\begin{center}
\includegraphics[scale=0.65]{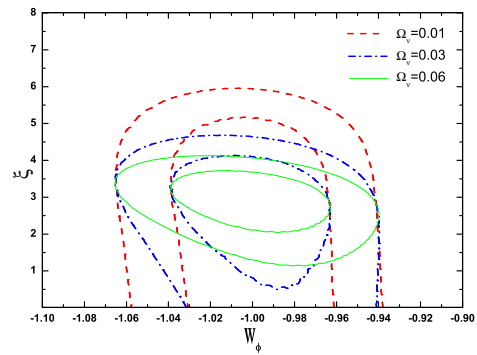} 
\caption{ Simulated 3-year SNAP constraints on the coupled
neutrino-DE model with neutrino energy density evolving as
$\rho_{\nu}=\frac{\Omega_{\nu 0}}{\Omega_{\phi
0}}\rho_{\phi}a^{-\xi}$, where the subscript '0' stands for today's
value. The red(dashed), blue(dash dotted) and green(solid) 2$\sigma$
contours stand for $\Omega_{\nu}=0.1 \%, 0.01,0.03$ and 0.06
respectively. The underlying fiducial model has been chosen as the
uncoupled $\Lambda$CDM model with
$\xi=3,w_{\phi}=-1,\Omega_{\phi}=0.7$ and $h=0.7$. We take a
gaussian prior $\sigma_m=0.01$. \label{fig:SNAPLC} }
\end{center}
\end{figure}

\section{Discussions and conclusion}

In this paper we have made the first study on the constraints on the
couplings between neutrinos and the scalar field of dark energy from
current and future observations of SNIa. We have shown that the
coupled system of neutrino dark energy can resemble the behavior of
Quintom and is fully degenerate with Quintom in light of the
geometric observations of SNIa only. A system with the phantom-like
dark energy coupled to massive neutrino is mildly favored by current
SNIa data where the neutrino mass is increasing with the redshift.
Such couplings are promising to be detected by future observations.

\section{acknowledgements}
Our simulations are computed in the Shanghai Supercomputer Center
(SSC). We thank Pei-Hong Gu for helpful discussions and Jean-Marc
Virey and  A. Tilquin for helpful comments. This work is supported
in part by National Natural Science Foundation of China (grant Nos.
90303004, 19925523) and by Ministry of Science and Technology of
China ( under Grant No. NKBRSF G19990754).

\end{document}